\documentclass[]{raa}
\usepackage{deluxetable}
\usepackage{amssymb}            
\usepackage{graphicx,times}
\usepackage{natbib}
\usepackage{color}

\providecommand{\bm}[1]{\mbox{\bol
dmath$#1$\unboldmath}}

\begin{document}

\title{Optical light curve of GRB 121011A: a textbook for the onset of GRB  afterglow in a mixture of ISM and wind-type medium}


\volnopage{}
        \setcounter{page}{1}

        \author{Li-Ping Xin
        \and Jian-Yan Wei
                \and Yu-Lei Qiu
                \and Jin-Song Deng
                \and Jing Wang
                \and Xu-Hui Han
        }
 \institute{Key Laboratory of Space Astronomy and Technology, National Astronomical Observatories, Chinese Academy of Sciences,
             Beijing 100012, China; {xlp@nao.cas.cn}
%
%
}

\abstract{
We reported the optical observations of GRB 121011A by 0.8-m TNT telescope at Xinglong observatory, China. The  light curve of optical afterglow  shows a smooth and featureless bump  during the epoch of $\sim$130 sec and $\sim$5000 sec with a rising index of $1.57\pm0.28$ before the  break time of $539\pm44$ sec, and a decaying index of about $1.29\pm0.07$ up to the end of our observations. Meanwhile, the X-ray light curve  decays in a single power-law with a slop of  about $1.51\pm0.03$ observed by $XRT$ onboard ${\rm} Swift$ from 100 sec  to about 10000 sec after the burst trigger. The featureless optical light curve could be understood as an onset process under the external-shock model.
The typical frequency has been below or near the optical one before the deceleration time, and the cooling
frequency is located between the optical and X-ray wavelengths.
The external medium density has a transition from a  mixed stage of ISM and wind-type medium before the peak time to the ISM at the later phase.
The joint-analysis of X-ray and optical light curves shows that the emission from both frequencies are consistent with the prediction of the standard afterglow model without any energy injections, indicating that the central engine has stopped its activity and does not restart anymore after the prompt phase.
\keywords{Gamma-ray bursts --- stars: individual (GRB 121011A)} }

 \authorrunning{Xin et al. }            
 \titlerunning{GRB 121011A }  
 \maketitle

\section{Introduction}

Gamma-ray burst (GRB;see e.g., Piran 2005, for a review) is the brightest flash in the universe.
Based on duration of prompt emission in high energy, GRBs are classified into two groups: long- and short- GRBs.
Long-duration GRBs are thought to be related to the death of massive stars ( e.g., Xin et al., 2011). The circum-burst medium density near the location of long-duration burst might be wind-type (e.g., Dai \& Lu 1998; Chevalier \&
Li 2000; Panaitescu \& Kumar
2002; Starling et al. 2008).
Multi-wavelength afterglows could be detected for almost half of GRBs when relativistic jet sweeps upon the environment medium.
Comparisons of optical and X-ray light curves show that their evolutions are usually different (Panaitescu et al. 2006; Huang et al. 2007a; Xin et al., 2012; De Pasquale et al., 2015). For some bursts like the naked-eye burst GRB 080319B ,they are likely to be generated by different outflows (Racusin et al., 2008).
Optical light curves show a diverse behavior at the early phase (Kann et al., 2011), which could be produced by several radiations, such as those related to the prompt emission, reverse-shock, forward shock, etc.
The brightness and the behavior in optical band could be a co-contribution by all these mechanisms or only dominated by one of them.
If there is only forward shock emission,
the brightness in low frequencies (e.g., optical, infrared, radio) is predicted to be increased over time at the first phase and then decreased at the later epoch after enough medium is swept by forward shock ( Sari \& Piran 1999). This kind of process is named as onset of the afterglow
which usually has a peak time of about 500 sec post the burst (Liang et al., 2013).

The onset process were reported in some bursts  (e.g., Molinari et al., 2007; Huang et al., 2012) , thanks to the  fast slew capability of $Swift$ satellite and  quick multi-wavelength follow-up observations by ground-based telescopes.
In those reports, other radiations from reverse shock or internal shock or others
were detected  in optical or X-ray light curves. For example, the variation in the optical light curve of GRB 060607 (Molinari et al., 2007) at the later phase was detected. During X-ray observations, flares (e.g., Burrows et al., 2005)  or emission in steep decay phase which is related with internal-shock process were detected  (e.g., Zhang 2007; Huang et al., 2012).
Recently, Li et al.,(2015) found that in 87 well sampled and simultaneously observed GRBs, only 9 GRBs
 match the prediction of the standard fireball model, and have a single
power law decay in both energy bands during their entire observations.
Some relationships are also investigated among different physical parameters like
the isotropic energy, peak time etc. (e.g., Liang et al., 2010; Panaitescu et al., 2011;2013).

In this work, we report a new ideal case of onset process in optical afterglows of GRB 121011A which was observed by 0.8m TNT telescope,
located at Xinglong observatory, China. Both optical and X-ray light curves during our observation epoch have no any pollution from other radiations. Thus it is a "purest" case for onset process.

\section{High-Energy Observation and data reduction}

At 11:15:30 UT, the Swift Burst Alert Telescope (BAT) triggered and
located GRB 121011A.
The light curve of prompt emission shows a single peak starting at -9 sec and ending at about 150 sec after the burst trigger.
The burst duration T$_{90}$ (15-350 keV) is $75.6\pm12.7$ sec (Stamatikos et al. 2012).
The XRT and UVOT began observing the burst at 97.5 sec and 150 sec after the burst, respectively, and the bright counterpart was found.
Follow-up observations were made by several ground-based telescopes, like MITSuME 50cm telescope (Kuroda et al., 2012, GCN 13846) and MASTER
(Yurkov et al. 2012, GCN 13848).

The XRT light curve and spectral data were obtained from the XRT light curve and spectral repository (Evans et al. 2007, 2009).
The XRT spectrum has been regrouped to ensure at least 3 counts
per bin using the grppha task from the XSPEC12.

\subsection{TNT observations}

Follow-up observation campaign of GRB 121011A was carried out using
the TNT  (0.8-m Tsinghua University - National Astronomical Observatory of China
Telescope) at Xinglong Observatory.
A PI $1300\times1340$ CCD and filters in the standard Johnson
Bessel system are equipped for TNT.
The observation of the optical transient of GRB 121011A was carried out with
TNT beginning at 114 seconds post the {\em Swift}/BAT trigger and
$W$ and $R$-band images were obtained.

Data reduction was carried out following the standard routine in
IRAF\footnote{IRAF is distributed by NOAO, which is operated by AURA, Inc., under
cooperative agreement with NSF.} package, including bias and flat-field
corrections. Dark correction was not performed since the temperature of the used CCD
was cooled down to $-110\,^{\circ}\mathrm{C}$. Point spread function (PSF)
photometry was applied via the DAOPHOT task in the IRAF package to obtain the
instrumental magnitudes. During the reduction, some frames were combined in
order to increase the signal-to-noise ratio ($S/N$).
Absolute calibration was performed using several nearby stars in the USNO B1.0 R2mag.
The data of GRB 121011A obtained in this work are reported in Table.1.

\begin{table}[!hc]
\caption{Optical Afterglow Photometry Log of GRB 121011A by TNT.
The reference time $T_0$ is {\em Swift} BAT burst trigger time.
"T-T0" is the middle time in second.
"Exposure" is the exposure time in second.
"Merr" means the uncertainty of magnitude.
All data are calibrated by nearby $USNO B1.0$ reference stars in R2 mag.
All Data are not
corrected for the Galactic extinction (which is $E_{B-V}=0.03$, Schlegel et al.1998).}
\centering
\begin{tabular}{lcccl}
\hline\hline
T-T0 & Exposure & Filter & Mag & Merr    \\
\hline
145   &  60   &  W  &  18.544  &  0.131  \\
215   &  60   &  W  &  17.908  &  0.056  \\
274   &  40   &  W  &  17.582   &  0.093  \\
330   &  60   &  W  &  17.490    &  0.043  \\
390   &  40   &  W  &  17.122   &  0.042  \\
416   &  20   &  W  &  17.057  &  0.093  \\
462   &  20   &  W  &  16.975  &  0.043  \\
485   &  20   &  W  &  17.040    &  0.042  \\
508   &  20   &  W  &  16.951  &  0.037  \\
551   &  40   &  W  &  17.014  &  0.029  \\
587   &  60   &  R  &  16.988  &  0.038  \\
666   &  60   &  R  &  17.019  &  0.034  \\
746   &  60   &  R  &  17.127  &  0.035  \\
824   &  60   &  R  &  17.222   &  0.040  \\
903   &  60   &  R  &  17.214   &  0.039  \\
982   &  60   &  R  &  17.315  &  0.042  \\
1061  &  60   &  R  &  17.427  &  0.055  \\
1218  &  60   &  R  &  17.445  &  0.050  \\
1297  &  60   &  R  &  17.612   &  0.051  \\
1376  &  60   &  R  &  17.565  &  0.062  \\
1455  &  60   &  R  &  17.806  &  0.062  \\
1533  &  60   &  R  &  17.913  &  0.087  \\
1612  &  60   &  R  &  17.866  &  0.071  \\
1691  &  60   &  R  &  17.905  &  0.065  \\
1770  &  60   &  R  &  17.970  &  0.082  \\
1848  &  60   &  R  &  17.803   &  0.075  \\
1927  &  60   &  R  &  18.238  &  0.124  \\
2006  &  60   &  R  &  18.052  &  0.084  \\
2085  &  60   &  R  &  18.331  &  0.116  \\
2165  &  300  &  R  &  18.416  &  0.061  \\
2483  &  300  &  R  &  18.506   &  0.068  \\
3119  &  300  &  R  &  18.976  &  0.102  \\
3437  &  300  &  R  &  19.276  &  0.133  \\
3755  &  300  &  R  &  19.037  &  0.127  \\
4073  &  300  &  R  &  19.227  &  0.145  \\
4391  &  300  &  R  &  19.242  &  0.133  \\
4709  &  300  &  R  &  19.581  &  0.213  \\
5027  &  300  &  R  &  19.397  &  0.225  \\
5345  &  300  &  R  &  19.367  &  0.314  \\
 \hline
\end{tabular}
\label{Tab3}
\end{table}

\section{Multi-band afterglows in temporal and spectral energy distribution}

Fig.1 shows the optical and X-ray light curves for GRB 121011A.
The reports from  GCN Circular (Fynbo et al., GCN 13856; Malesani et al., GCN 13853; )
after the TNT observation epoch are collected in order to get larger coverage of light curves.
All these magnitudes reported  were re-calibrated by the same stars with USNO B1.0 R2 mag.
 The well sampled R-band light curve in this work
during the epoch of $\sim100$ sec and 5400 sec post burst
was well fitted by a broken  power-law
\begin{equation} F=F_0\left [
\left (   \frac{t}{t_b}\right)^{\omega\alpha_1}+\left (
\frac{t}{t_b}\right)^{\omega\alpha_2}\right]^{-1/\omega}.
\end{equation}

The temporal decaying indices are changed from  $\alpha_{O1}=-1.57\pm0.28$ to
$\alpha_{O2}=1.29\pm0.07$ with a break time  $t_b$ of $539\pm44$ sec and a smoothness parameter $w$ of $1.13\pm0.34$. The $\chi^2$ is 50.4 with a degree of freedom of 36. The smoothness parameter $w$  is similar with the results of
GRB 060604 and GRB 060607A (Molinari et al. 2007). We also note that
these values are smaller than those of  most bursts (Liang et al. 2010,2013).

During the similar observation epoch as optical one,
X-ray light curve could be fit well with a single power law  $f \sim t^{-\alpha}$ with a
temporal decay index of $\alpha_{x1}=1.51\pm 0.03$ .
The $\chi^2$  is 26.8 with a degree of freedom of 20.

Fig.2 shows the spectral energy distribution from optical to X-ray afterglows of GRB 121011A during 3000 and 4000 sec post burst. The spectral index of X-ray was obtained $\beta_{x}=-1.1\pm0.65$ after fitting the data with a model of a single power law ($f_{\nu}\sim\nu^{\beta}$) plus absorption due to the ISM including molecules and grains . The uncertainty of $\beta_x$ is relatively large. This is because the data in the fitting region is sparse. However, if regardless of the uncertainty. This optical data is located below the
extrapolation of the X-ray spectral emission.

\begin{figure}[!hc]
\centering
\includegraphics[angle=-90,scale=0.5]{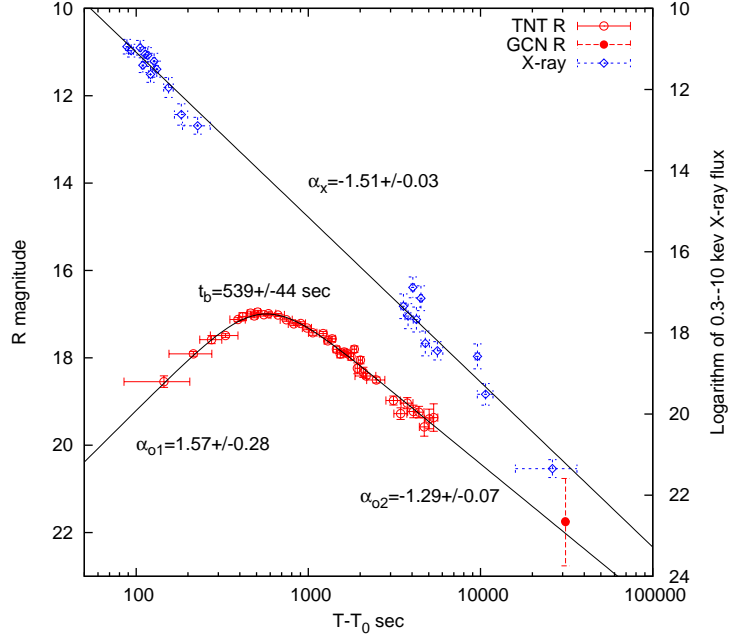}
\begin{flushleft}
\caption{ The optical and X-ray afterglow light curves of GRB 121011A.  Red data shows the R-band light curve observed by TNT telescope, and is calibrated by USNO B1.0 R2 mag.  Blue data shows the X-ray light curve. All X-ray data is transformed with the formation of -12-2.5$\times$$\log_{10}f$ for comparison clearly.}
\end{flushleft}
\label{Fig1}
\end{figure}

\begin{figure}[!hc]
\centering
\includegraphics[angle=-90,scale=0.5]{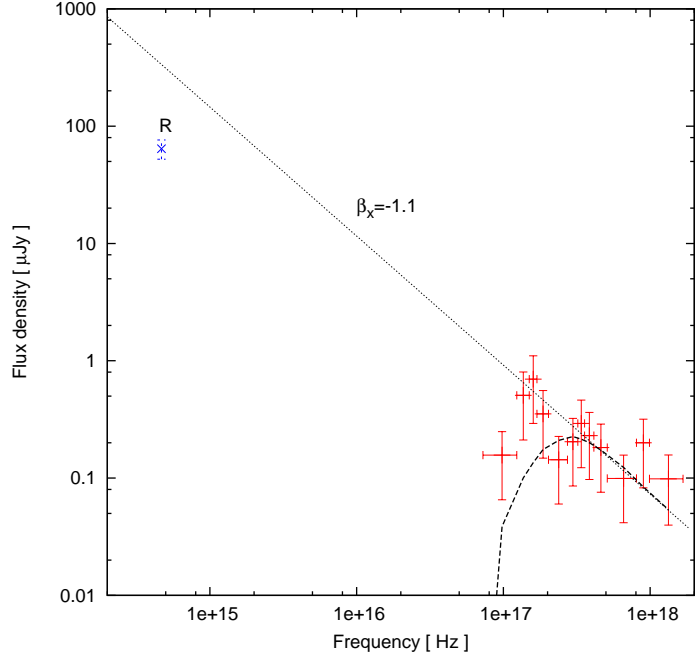}
\begin{flushleft}
\caption{Optical to X-ray spectral energy distribution of GRB 121011A during 3000 and 4000 sec post burst. The optical one is corrected from the Galactic extinction. The black dash line shows the best fit for the X-ray emission, and the black dot line shows the extrapolation for the spectral index ($f_\nu\sim\nu^\beta$) of X-ray emission.
}
\end{flushleft}
\label{Fig1}
\end{figure}

\section{Analysis and Discussion}
\subsection{Estimate the redshift }
Redshift is one of essential parameters for constraining the properties of a burst. However, there is no exact report of the redshift for GRB 121011A in the literature.
For the following analysis, we try to restrict the redshift based on the multi-wavelength observations in the literature, especially the blue band ones.
According to the reports by Holland et al. (2012),
during the similar epoch observed by $UVOT$ onboard $Swift$ ,
a  successful detection of about $19.67\pm0.07$ mag was obtained in blue band $u$,
while upper limits of 20.9, 22.6 and 21.0 mag were only given by more bluer filter of $uvw1,uvm2,uvw2$ respectively.
Assuming that the failure of afterglow detection for $uvw1$ band is caused by $Ly\alpha$ forest, while $u$ filter is not,
the $Ly\alpha$ absorption line would be estimated in the region of $2600<1216(1+z)<3465$, which conversely  makes the redshift to be in the region of [1.13, 1.84], considering the central wavelengths of $u$ (3486$\AA$) and $uvw1$ (2600$\AA$) respectively (Roming et al., 2005).
For simplicity, the redshift of 1.5 is adopted in our discussion in the next sections,
 which does not affect much of the main conclusions in this work.

\subsection{Test with the standard forward shock model}

An interesting signature for the afterglow of GRB 121011A is that
the chromatic behavior during the early phase in which optical one is rising while X-ray emission has been in the deceleration phase.
According to the forward shock synchrotron model,
electrons accelerated in the shock emit
synchrotron radiation.
As the fireball slows down, the synchrotron peak frequency moves progressively to a lower
frequency.
The optical light curve of the forward-shock emission is expected to show an initial rising with
an index of -0.5  to the peak, and then followed by a normal decay ($\alpha \sim 1$).
For the case of GRB 121011A, the index in the
rising phase is larger than 0.5 (Zhang et al., 2003, Wang et al., 2008),
 and no any rebrightening exists in the following decay phase. All these properties indicate that
the typical frequency  has moved to optical band ($\nu_m \le \nu_o$) before the peak time observed in light curve
(Type III case in Jin \& Fan 2007)
of about $\sim 539$ sec.

For the case of GRB 121011A, the decay index of optical afterglow after 539 sec post burst
is $1.29\pm0.07$, while that of X-ray light curve is $1.51\pm0.03$ during the same epoch.
The difference of decay indices between optical and X-ray afterglow is about 0.22, which is
consistent with the case of $\nu_o < \nu_c < \nu_X$ in the ISM under slow cooling and without any injection in the frame of the standard forward shock model (Zhang et al. 2006; Gao et al., 2013),where
$\nu_c$ is cooling frequency of synchrotron radiation.
This scenario is also consistent with the result in Fig.2 that optical flux is smaller than the prediction of X-ray emission, if  the uncertainty of the spectral slop $\beta_x$ were regardless.
If this is the case, the electrical energy index $p$  would be estimated to be about   2.68 with the relationship of $p=(4\times\alpha_x +2)/3$.

\subsection{Constrain the circumburst medium before the deceleration phase }
Long-duration GRBs are thought to be related to the deaths of massive stars,
which indicates that the environment around the bursts would not be  ISM but a wind-type-like medium.
Within the standard forward shock model, the slopes in rising and decaying phase of an afterglow onset depend on the circumburst medium density profile parameters $k$ in the profile
of $n=n_0(R/R_t)^{-k}$ (Liang et al., 2013),
 where $n_0=1$ cm$^{-3}$,$R_t$ is the transition radius at which the medium transits into
 a constant density ISM from wind-type medium.
In the spectral regime of $\nu_m <\nu_o < \nu_c$ and $p>2$,
the rising slope $\alpha$ is predicted to be a constant value of 3 for the ISM case  (e.g., Sari et al., 1999; Huang et al., 2007b), and it will be a varying value with the relation of $\alpha=(p-1)/2$
for the wind model.
With the estimation of $p\sim2.68$, the $\alpha$ would be deduced to be $(2.68-1)/2=0.84$
. The predicted ones in  both ISM and wind cases are not  consistent with  the rising one for the case of GRB 121011A.

Considering the circumburst medium density profile parameter $k$
in the thin-shell forward-shock model (Liang et al., 2013) , it gives a rising index $\alpha=-3+k(p+5)/4$.
For GRB 121011A, the parameter $k$ is deduced to be about 1.3
with the values of $\alpha=-1.51$ and $p=2.68$ which is estimated above by decaying part
analysis.
This result show that at the early epoch before the deceleration time, the medium density
is the mixture of ISM and wind-type medium.

\subsection{No any signatures   of reverse shock and long lasting central engine activity}

According to the standard relativistic fireball model,  reverse shock are expected to radiate
emissions in the long wavelength bands in optical, infrared and radio with a short bright optical flashes(e.g., Akerlof et al. 1999) and/or intense radio afterglows(e.g., Kulkarni et al. 1999) by executing
a synchrotron process in a particularly
early afterglow phase (e.g., Kobayashi 2000).
The non-detection of reverse-shock in optical band in GRB 121101A could be classified as the type-III case (Jin \& Fan 2007).
The lack of optical flash is naturally explained in the
standard model if the typical frequency of the forward shock
emission is lower than optical band $\nu_m < \nu_o$ at the onset of
afterglow, as discussed in the above section. At the peak time,
the forward and reverse shocked regions have the same
Lorentz factor and internal energy density. The cooling frequency
of the reverse shock is equal to that of the forward shock.
 The matter density in the reverse shocked region is much
higher than in the forward shock region, making the electron
temperature lower. The typical frequency of the reverse shock therefore is
much lower than that of the forward shock (e.g.,  Kobayashi \& Zhang 2003; Mundell et al., 2007; Melandri et al. 2010).
Other possibility for the lack of optical flash is that all the emission is produced in
a magnetically dominated outflow(Jin \& Fan 2007; Zhang et al., 2003).

In the $Swift$ era,  the behavior of X-ray emission at early phase is very complex. High latitude prompt emission (the first steep decay phase) and flares in the early epoch are observed in most of bursts( e.g., Chincarini et al., 2010), which is thought to be related to the central engine activity. Shallow-decay phase (or plateau phase) in the canonical light curve (e.g., Zhang et al., 2006) is also considered to be caused by energy injection (Dai \& Lu 1998,2000; Fan \& Xu 2006; Yu \& Huang 2013) from long-lasting central engine activity or others (for the review, see Zhang 2007).
Unlike most bursts reported before,
GRB 121101A shows a single decay during its whole X-ray observations. No any steep decay phase or flares or shallow-decay phase is observed as early as the beginning observation of $t\sim100$ sec,
indicating that the central engine stops its activity
at $\sim100$ sec post its initial burst.  The emission related to the prompt emission must
decay very rapidly from a very early time.
Furthermore, the central engine does not restart anymore at the later phase.

\subsection{Onset process and constrain of initial Lorentz factor }

For the smooth and featureless light curve in the optical afterglow of GRB 121011A, it is clearly
consistent with the prediction of onset process when the relativistic
jet meet the environment medium around the burst(e.g., Sari \& Piran 1999; Kobayashi \& Zhang 2007).
Since the break time of about 539 sec post burst from rise part to decay one
in the optical light curve, the peak time would be inferred to be $t_{peak}=t_b(-\alpha_{o1}/\alpha_{o2})^{1/[w(\alpha_{o2}-\alpha_{o1})]}\sim572\pm50$ sec
following the estimation of Molinari et al. (2007).
The peak time $t_p$ is larger than the duration of high-energy prompt emission of about $100$ sec, making the burst to be a thin-shell case.
In this scenario, the quantity $t_{peak}/(1+z)$ corresponds to the deceleration timescale $t_{dec}\sim R_{dec}/(2c\Gamma_{dec}^2)$, where $R_{dec}$ is the deceleration radius, $c$ is the speed of light and $\Gamma_{dec}$ is the fireball Lorentz factor at $t_{dec}$.
Therefore, initial Lorentz factor $\Gamma_{0}$ could be estimated since it is twice of $\Gamma_{dec}$ (Panaitescu \& Kumar 2000).

If the environment were a homogeneous medium, the initial Lorentz factor could be estimated with the relation
$\Gamma_0\sim193(n\eta)^{-1/8}\times(E_{\gamma,iso,52}/t_{p,z,2}^3)^{1/8}$ (Liang et al., 2010).
According to the joint-analysis of
X-ray to optical afterglows of GRB 1201011A in above sections, the burst environment does not fully agree with ISM,
it has a distribution with a parameters of $k\sim1.3$.
If  considering the case of  $R \sim R_{dec}$,
the external medium density is $n \sim n_0 \sim 1$. Under these conditions,
the estimation above for initial
Lorentz factor could be adopted, and $\Gamma_0$ is less dependent on $n$, $\eta$ and $E_{\gamma,iso,52}$.
As a result, a relationship between initial Lorentz factor and peak time in the
rest-frame was deduced by Liang et al. (2010),
which was $log\Gamma_0 =(3.59\pm0.11)-(0.59\pm0.05)log t_{p,z}$.
For the case of GRB 121011A, based on the relationship,
the redshift is roughly about 1.5,
peak time would be $t_{p,z}\sim 230$.
If it were the case, $\Gamma_0$ is inferred to be 157.

\section{Summary}

The optical light curve of GRB 121011A shows a  featureless bump  during the epoch of $\sim$130 sec and $\sim$5000 sec with a rising index of $1.57\pm0.28$ before the  break time of $539\pm44$ sec, and a decaying index of about $1.29\pm0.07$ up to the end of our observations.
 Meanwhile, the X-ray light curve decays in a single power-law with a slop of  about $1.51\pm0.03$ from 100 sec  to about 10000 sec after the burst.
 The joint-analysis of X-ray and optical light curves  after the peak time shows that the emission from both frequencies is consistent with the prediction of the standard afterglow model under the condition of ISM, and cooling frequency is located between optical and X-ray ones.
At the early phase, the rising index in optical light curve is different from the predictions of the standard model under the cases of ISM and wind-type, but is consistent with the condition that the external medium is at a mixed stage of ISM and wind-type one.
Besides, the typical frequency $\nu_m$ has moved to the optical one before the deceleration time, since there is no signature for the transition by $\nu_m$ from optical frequency (Jin \& Fan 2007).
No any radiation from other  mechanisms like reverse shock or those related to the internal shock,
are detected in both optical and X-ray light curves during all the observation epoch,
making this burst as a "purest" one for the onset of afterglows.

\begin{acknowledgements}
This work was supported by the National Basic
Research Program of China (973 program, 2014CB845800) and the National Natural Science
Foundation of China (Grant No. U1331202).
This work made use of data supplied by the UK Swift Science Data Centre at the University of Leicester.
XLP acknowledges the support of the National Natural Science Foundation
of China (NSFC) grant 11103036 and U1331101.
Y. Qiu is supported by the National Natural Science Foundation of China (No. U1231115).
\end{acknowledgements}

\label{lastpage}

\end{document}